\documentclass[twocolumn,pra,superscriptaddress]{revtex4}
\usepackage[utf8]{inputenc}
\usepackage{amsmath}
\usepackage{amssymb}
\usepackage{graphicx}
\usepackage{esint}

\makeatletter
\@ifundefined{textcolor}{}
{%
 \definecolor{BLACK}{gray}{0}
 \definecolor{WHITE}{gray}{1}
 \definecolor{RED}{rgb}{1,0,0}
 \definecolor{GREEN}{rgb}{0,1,0}
 \definecolor{BLUE}{rgb}{0,0,1}
 \definecolor{CYAN}{cmyk}{1,0,0,0}
 \definecolor{MAGENTA}{cmyk}{0,1,0,0}
 \definecolor{YELLOW}{cmyk}{0,0,1,0}
}

\usepackage{subfigure}\usepackage{epsfig}\usepackage{amsfonts}\usepackage{mathrsfs}\usepackage{CJK}\usepackage[toc,page,title,titletoc,header]{appendix}

\makeatother
\begin{document}
\title{Signatures of Bose-Einstein condensation in an optical lattice}

\author{Ke-Ji Chen}
\affiliation{Department of Physics, Renmin University of China, Beijing, 100872,
China}
\author{Jingkun Wang}
\affiliation{Department of Physics, Renmin University of China, Beijing, 100872,
China}
\author{Wei Yi}
\email{wyiz@ustc.edu.cn}
\affiliation{Key Laboratory of Quantum Information, University of Science and Technology of China,
CAS, Hefei, Anhui, 230026, China}
\affiliation{Synergetic Innovation Center of Quantum Information and Quantum Physics, University of Science and Technology of China, Hefei, Anhui 230026, China}
\author{Wei Zhang}
\email{wzhangl@ruc.edu.cn}
\affiliation{Department of Physics, Renmin University of China, Beijing, 100872,
China}
\affiliation{Beijing Key Laboratory of Opto-electronic Functional Materials and Micro-nano Devices,
Renmin University of China, Beijing 100872, China}

%
\begin{abstract}
We discuss typical experimental signatures for the Bose-Einstein condensation (BEC) of an ultracold Bose
gas in an inhomogeneous optical lattice at finite temperature. Applying the
Hartree-Fock-Bogoliubov-Popov formalism, we calculate quantities such as the
momentum-space density distribution, visibility and
peak width as the system is tuned through the superfluid to normal phase transition. Different from
previous studies, we consider systems with fixed total particle number, which is of direct experimental
relevance. We show that the onset of BEC is accompanied by sharp features in all these signatures, which
can be probed via typical time-of-flight imaging techniques. In particular, we find a two-platform
structure in the peak width across the phase transition. We show that the onset of condensation is related
to the emergence of the higher platform, which can be used as an effective experimental signature.
\end{abstract}
\maketitle

\section{Introduction}
\label{intro}

The realization of Bose-Hubbard model in ultracold atomic gases and the subsequent observation of the
superfluid to Mott-insulator phase transition represents a milestone for the quantum simulation of strongly
correlated many-body systems in ultracold atomic gases~\cite{jaksch-98,greiner-02,spielman-08,mark-11,deMarco-11}.
In these experiments, the onset of the superfluid phase is typically connected with the emergence of
interference peaks in the time-of-flight images. For example, in Ref.~\cite{greiner-02}, high visibility of
interference peaks is taken as an indicator for the existence of superfluidity, while in
Ref.~\cite{spielman-08} and \cite{mark-11}, the onset of a bimodal distribution and
the rising point of the peak width of the interference pattern in the first Brillouin zone are used
as the signature, respectively.

The interference pattern in the post expansion image originates from the existence of short-range correlations.
In a uniform non-interacting Bose gas, the short-range correlations can even persist well above the transition
temperature, leading to the presence of interference pattern~\cite{zhai-07}. However, more careful
calculations including inter-atomic interactions and global trapping potential confirm a sharp change
in visibility at the phase transition point, hence validate the usage of interference pattern as a
signature~\cite{yi-07,lin-07,gerbier-08}. In Ref.~\cite{yi-07,lin-07}, it is further suggested that a
bimodal structure and the sharp change of the interference peak width in the first Brillouin zone can
also serve as an unambiguous signature of superfluidity. In both of these studies, the chemical potentials
at the center of the global trapping potential have been fixed, leading to a situation where the total
number of particles is not fixed as the temperature is tuned across the thermal phase transition
between the superfluid state and the normal state. This scheme is in clear contrast to existing experiments,
where the superfluid to normal phase transition is usually tuned through by varying the optical lattice
depth with a fixed total particle number~\cite{greiner-02,spielman-08,mark-11}. To date, a detailed
finite-temperature characterization of these experimental signatures for a system with fixed total
number of particles is still lacking.

In this work, as an extension to previous studies, we investigate the finite-temperature properties of a trapped
ultracold Bose gas throughout the superfluid to normal phase transition, while the total particle number
of the system is kept fixed. We apply the Hartree-Fock-Bogoliubov-Popov (HFBP)
formalism~\cite{shi-98,vanoosten-01,rey-03,andersen-04,blakie-09-a,blakie-09-b},
and focus on the characterization of the commonly used signatures of lattice superfluidity,
including the visibility, the bimodal structure and the peak width of the interference pattern.
We find that all these quantities demonstrate pronounced features as the system crosses the
critical point, which can be measured in typical time-of-flight imaging experiments. Interestingly,
we demonstrate that as the lattice depth is tuned through the critical depth, a two-platform structure
can show up in the peak width measurement of the interference peak in the first Brillouin zone.
Based on the results of our calculation, we propose to detect the critical temperature using the
second platform as an unambiguous signature.

This paper is organized as the following: in Sec. \ref{model}, we present the HFBP formalism.
In Sec. \ref{results}, we apply the local density approximation and the HFBP formalism to characterize
the momentum distribution of the trapped gas across the critical temperature. We identify sharp features
in all three signatures close to the critical temperature for a system with fixed total number of particles.
For the peak width in the first Brillouin zone, we propose to associate the critical temperature with the
appearance of a second platform. Finally, we summarize in Sec. \ref{summary}.

\section{Formalism}
\label{model}

In this section, we present the HFBP formalism, which, combined with the local density approximation, is used to characterize a trapped BEC at finite temperatures. The Hamiltonian of our system can be written as:
\begin{eqnarray}\label{Hamiltonian}
  H &=& \int d^{3}\textbf{r}\Psi^{\dag}(\textbf{r})\Big[-\frac{\hbar^{2}}{2m}\nabla^{2}+V_{\textrm{op}}(\textbf{r})+V(\textbf{r})\Big]\Psi(\textbf{r})\nonumber\\
  &&+\frac{U_{\rm bg}}{2}\int d^{3}\textbf{r}\Psi^{\dag}(\textbf{r})\Psi^{\dag}(\textbf{r})\Psi(\textbf{r})\Psi(\textbf{r}).
\end{eqnarray}
where the lattice potential $V_{\textrm{op}}(\textbf{r})\equiv V_{0}\sum_{i=x,y,z}\sin^{2}(\pi r_{i} /d)$ with $d$ the
lattice spacing and $V_{0}$ the lattice depth, the global harmonic trapping potential $V(\textbf{r})\equiv m \omega^{2}
r^{2} /2$, and the $s$-wave interaction rate $U_{\textrm{bg}}=4\pi \hbar^{2} a_{s}/m$. We consider the tight-binding
case, where the lattice potential is strong enough such that atoms are tightly confined in each lattice site.
As a comparison, the global trapping frequency is much weaker. Under these conditions,
we can employ the single-band approximation, where the Bose filed operators can be expanded in the basis of the Wannier functions $w(\textbf{r}-\textbf{r}_{i})$ of the lowest band $\Psi (\textbf{r})=\sum_{i}w(\textbf{r}-\textbf{r}_{i})a_{i}$ \cite{jaksch-98}. The Hamiltonian Eq.~(\ref{Hamiltonian}) can then be reduced to the Bose-Hubbard Hamiltonian:
\begin{eqnarray}\label{Bose_Hubbard model}
  H = -t\sum_{<i,j>}a^{\dag}_{i}a_{j}+\frac{U}{2}\sum_{i}a^{\dag}_{i}a^{\dag}_{i}a_{i}a_{i}
  -\sum_{i}{\mu_{i}}a^{\dag}_{i}a_{i},
\end{eqnarray}
where the local chemical potential $\mu_{i}=\mu(0)-V(\textbf{r}_{i})$ under
the local density approximation. In the remainder of this manuscript, we use the recoil energy
$E_{\textrm{R}}\equiv \hbar^{2}\pi^{2}/(2md^{2})$ as the unit of energy, and the lattice constant
$d$ as the unit of length. In this unit system, the dimensionless hopping rate
$t \approx (3.5/\sqrt{\pi})s^{3/4}\exp(-2\sqrt{s})$ with $s\equiv V_{0}$~\cite{duan-05}.
The on-site repulsive interaction rate $U$ is defined as
$U\equiv U_{\rm bg}\int |w(\textbf{r})|^{4}d\textbf{r}$. In typical experiments,
the dimensionless interaction $U \approx 3.05 s^{0.85}a_{s}$~\cite{gerbier-05}.

Under the local density approximation, the inhomogeneous Bose gas described by Hamiltonian
Eq. (\ref{Bose_Hubbard model}) can be considered as a group of uniform subsystems with
a slowly varying local chemical potential. Each subsystem can then be described by a uniform
Hamiltonian, which can be transformed into momentum space by introducing the creation and annihilation
operators $a^{\dag}_{\textbf{k}}$ and $a_{\textbf{k}}$:
\begin{eqnarray}
  a_{i} &=& \frac{1}{\sqrt{\cal V}}\sum_{\textbf{k}}a_{\textbf{k}}e^{-i \textbf{k} \cdot \textbf{r}_{i}} ,\nonumber \\
  a^{\dag}_{i} &=& \frac{1}{\sqrt{\cal V}}\sum_{\textbf{k}}a^{\dag}_{\textbf{k}}e^{i \textbf{k} \cdot \textbf{r}_{i}}.
  \label{c and c_dag}
\end{eqnarray}
Here, ${\cal V}$ is the dimensionless quantization volume, which is the number of lattice sites.
Substituting Eq.~(\ref{c and c_dag}) into the Bose-Hubbard Hamiltonian Eq.~(\ref{Bose_Hubbard model}),
we find:
\begin{eqnarray}
  H =&& \sum_{\textbf{k}} (\epsilon_{\textbf{k}}-\mu)a^{\dag}_{\textbf{k}}a_{\textbf{k}}\nonumber\\
  &&+\frac{U}{2 {\cal V}}\sum_{\textbf{k},\textbf{k}',\textbf{q}}a^{\dag}_{\textbf{k}+\textbf{q}}a^{\dag}_{-\textbf{k}}a_{\textbf{k}'+\textbf{q}}a_{-\textbf{k}'},
\end{eqnarray}
where  $\epsilon_{\textbf{k}}=-2t\sum_{i=x,y,z}\cos(k_{i}d)$ is the lattice dispersion in the lowest band.

Under the standard HFBP approach, we obtain the effective Hamiltonian \cite{vanoosten-01,rey-03,andersen-04}
\begin{eqnarray}\label{effective Hamiltonian}
  H_{\textrm{eff}} & \approx & (\epsilon_{0}-\mu  +\frac{U n_{0}}{2})N_{0}+\sum_{\textbf{k}\neq 0}[\epsilon_{\textbf{k}}-\mu+ 2Un_{\textrm{tot}}]a^{\dag}_{\textbf{k}}a_{\textbf{k}}\nonumber \\
  && +\frac{Un_{0}}{2}\sum_{\textbf{k} \neq 0}[a^{\dag}_{\textbf{k}}a^{\dag}_{-\textbf{k}}+a_{\textbf{k}}a_{-\textbf{k}}]
  \nonumber \\
  && -\frac{U}{\cal V}\sum_{\textbf{k},\textbf{k}'\neq 0}
  n_{\rm ex}({\bf k}) n_{\rm ex}({\bf k}^\prime),
\end{eqnarray}
where $N_{0}$ is the total particle number in the condensate with the condensate filling factor $n_{0}=N_{0}/{\cal V}$,
and $n_{\rm tot}$ is the total filling factor.
Here, $n_{\rm ex}({\bf k})=\langle a^{\dag}_{\textbf{k}}a_{\textbf{k}}\rangle$ for ${\bf k} \neq 0$
is the momentum space distribution of the thermal component,
and $\sum_{\textbf{k},\textbf{k}'\neq 0} n_{\rm ex}({\bf k}) n_{\rm ex}({\bf k}^\prime) =(N-N_{0})^{2}$.
Under HFBP, the chemical potential is given as:
\begin{eqnarray}
  \mu &=&  \frac{dE}{dN}\Big|_{S}=\frac{d \langle H_{\textrm{eff}}\rangle}{dN}\Big| _{S}.\label{chempot}
\end{eqnarray}

The effective Hamiltonian above can be diagonalized using the standard Bogoliubov transformation:
\begin{eqnarray}
  H_{\textrm{eff}} &=& (\epsilon_{0}-\mu  +\frac{U n_{0}}{2})N_{0}
  \nonumber \\
  && +\frac{1}{2}\sum_{\textbf{k}\neq 0}[\lambda-(\epsilon_{\textbf{k}}-\mu+2Un_{\textrm{tot}})]\nonumber \\
  && +\sum_{\textbf{k}\neq 0}\lambda \alpha^{\dag}_{\textbf{k}} \alpha_{\textbf{k}}-\frac{U}{\cal V}(N-N_{0})^{2},
\end{eqnarray}
where $\lambda=\sqrt{(\epsilon_{\textbf{k}}-\mu+2Un_{\textrm{tot}})^{2}-(Un_{0})^{2}}$ is the quasiparticle dispersion relation and  $\alpha^{\dag}_{\textbf{k}}$ ($\alpha_{\textbf{k}}$) is the creation (annihilation) operator for the Bogoliubov quasi-particles.

The thermodynamic potential at a finite temperature $T$ is given by $\Omega=-(1/\beta)\ln {\rm Tr}(e^{-\beta H_{\textrm{eff}}})$, where $\beta=1/k_BT$ with $k_B$ the Boltzmann constant.
The total filling factor can then be determined by the number equation, leading to
\begin{eqnarray}\label{number equation 1}
  n_{\textrm{tot}} &=& n_{0}+\frac{1}{2 {\cal V}} \sum_{\textbf{k}\neq 0}
  \left[\frac{\epsilon_{\textbf{k}}-\epsilon_{0}+Un_{0}}{E_{\textbf{k}}}
  \coth \left( \frac{\beta E_{\textbf{k}}}{2} \right)-1\right], \nonumber \\
\end{eqnarray}
where $E_{\textbf{k}}=\sqrt{(\epsilon_{\textbf{k}}-\epsilon_{0}+Un_{0})^{2}-(Un_{0})^{2}}$.
For a normal gas, the number equation can be obtained from Eq. (\ref{number equation 1})
by setting $n_0=0$. The total particle number of such a system is given by
\begin{eqnarray}\label{total number}
  N &=& \int n_{\textrm{tot}}(\textbf{r} ) d^{3}\textbf{r}.
\end{eqnarray}
We may then solve Eqs. (\ref{number equation 1}) and (\ref{total number}) self-consistently
to determine the chemical potential at trap center, the corresponding density and momentum
distributions for a given temperature $T$ and total particle number $N$.
\begin{figure}
\begin{center}
\includegraphics[width=8.0cm]{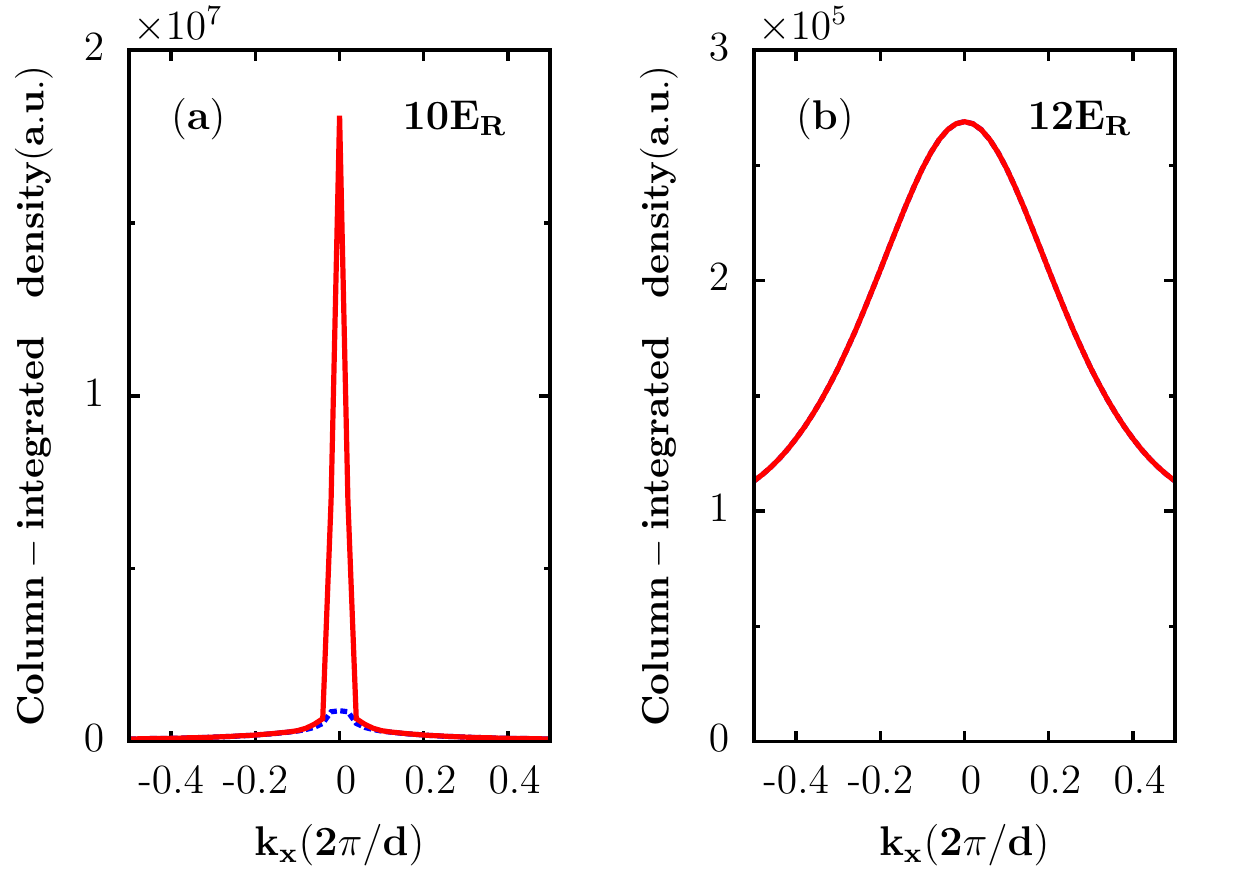}\\
\caption{(Color online) Momentum-space column-integrated density along the $x$-axis in the first Brillouin zone. A clear bimodal structure appears when $V_{0}$ is below  the critical value. Here, the solid curves (red) are the total momentum density distribution, and the dash curves (blue) are the momentum distribution of the thermal component. For the numerical calculations, the trapping frequency $\omega=2\pi \times 40$ Hz, the total number $N=1.0\times 10^{5}$, and the temperature $T=0.18E_{\textrm{R}}$. The lattice depth for the subplots are: (a) $V_{0}=10E_{\textrm{R}}$, (b) $V_{0}=12E_{\textrm{R}}$.}
\end{center}
\end{figure}

\section{Experimental signatures}\label{results}

In this section, we characterize various experimental signatures for the onset of superfluidity using HFBP formalism, which essentially relies on the calculation of momentum space distribution of the trapped lattice gas. The momentum space distribution can be obtained by transforming the field operator to the momentum space
\begin{eqnarray}
  \Psi(\textbf{k}) &=& w(\textbf{k})a_{\textbf{k}},
\end{eqnarray}
where $\Psi(\textbf{k})$, $w(\textbf{k})$ and $a_{\textbf{k}}$ are the Fourier components of
$\Psi(\textbf{r})$, $w(\textbf{r})$ and $a_{i}$, respectively.
As a result, the actual atomic  momentum distribution takes the form
\begin{eqnarray}
  n(\textbf{k}) &=& \langle \Psi^{\dag}(\textbf{k})\Psi(\textbf{k}) \rangle =|w(\textbf{k})|^{2}\langle a^{\dag}(\textbf{k})a(\textbf{k})\rangle\nonumber \\
  &=& |w(\textbf{k})|^{2}(n_{0}(\textbf{k})+n_{\textrm{ex}}(\textbf{k})),
 \end{eqnarray}
where $n_{0}(\textbf{k})$ is the momentum space distribution of the condensate.
While the momentum distribution of the thermal gas can be obtained from the trap integration:
\begin{eqnarray}
 && n_{\textrm{ex}}(\textbf{k}) =\nonumber\\
    && \hspace{5mm}
    \frac{1}{2{\cal V}} \int d^{3}\textbf{r} \left[\frac{\epsilon_{\textbf{k}}-\epsilon_{0}+Un_{0}}{E_{\textbf{k}}}\coth\frac{\beta E_{\textbf{k}}}{2}-1\right],
\end{eqnarray}
the condensate momentum distribution is obtained from the Fourier transformation of the condensate wave
function~\cite{dalfovo-99}
\begin{eqnarray}\label{psi_0}
  \psi_{0}(\textbf{k}) &=& \frac{1}{\sqrt{\cal V}}\int \psi_{0}(\textbf{r})e^{- i \textbf{k}\cdot \textbf{r}}d^{3}\textbf{r}
\end{eqnarray}
via the relation $n_{0}(\textbf{k})=|\psi_{0}(\textbf{k})|^{2}$.
\begin{figure}
\begin{center}
  \includegraphics[width=8.5cm]{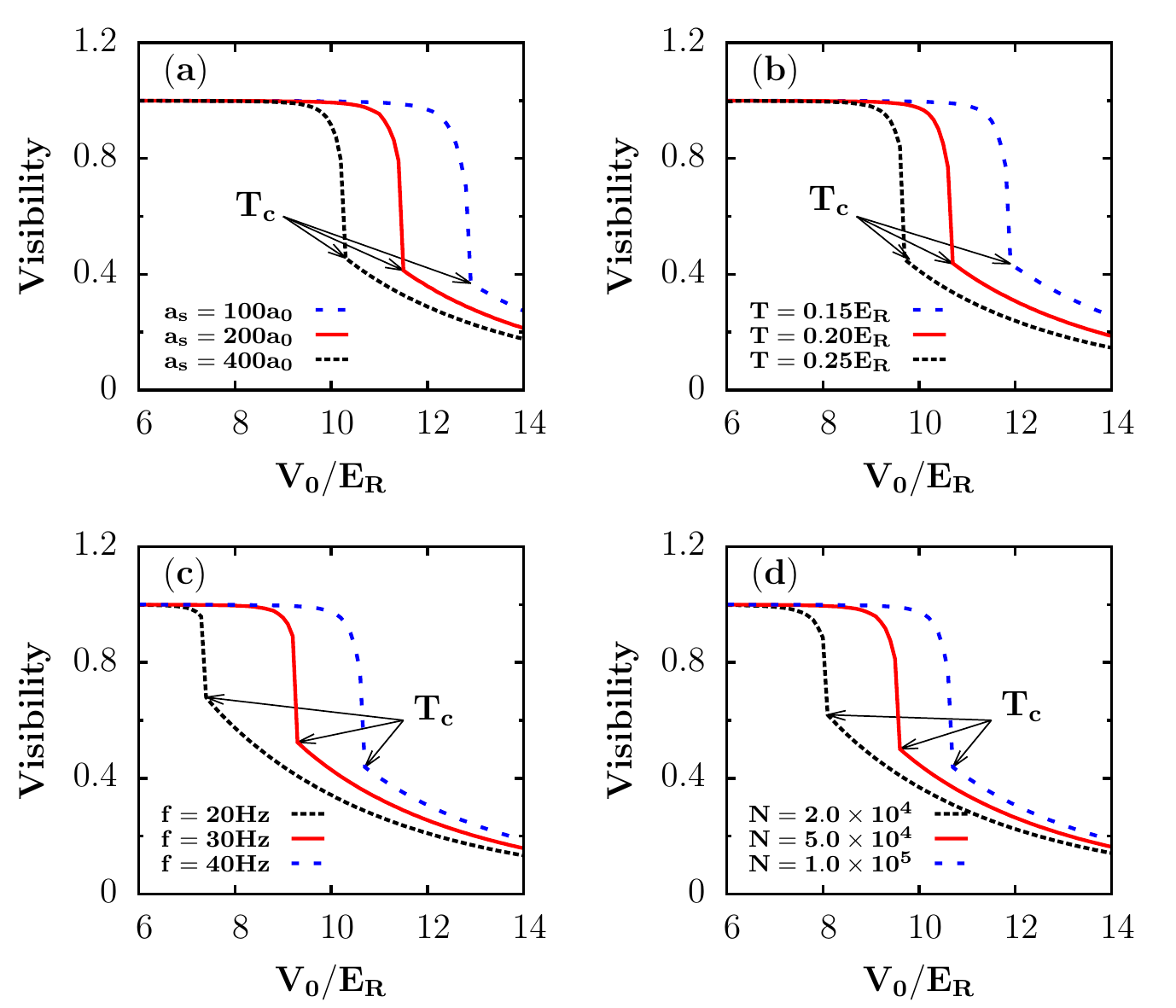}\\
\caption{(Color online) Visibility as a function of scattering length, temperature, trapping frequency and total particle number, respectively. The parameters are: (a) $T=0.2E_{\textrm{R}}$, $\omega=2\pi\times 40$ Hz and $N=1.0\times 10^{5}$; (b) $a_{s}=320a_{0}$, $\omega=2\pi\times 40$ Hz and $N=1.0\times 10^{5}$; (c) $a_{s}=320a_{0}$, $T=0.2E_{\textrm{R}}$ and $N=1.0\times 10^{5}$; (d) $a_{s}=320a_{0}$, $\omega=2\pi\times 40$ Hz and $T=0.2E_{\textrm{R}}$.}
\end{center}
\end{figure}

Experimentally, the momentum distribution of the trapped gas is typically obtained via a time-of-flight image,
which is essentially a column-integrated momentum distribution along the direction of the probe laser:
\begin{eqnarray}
  n_{\perp}(k_{x},k_{y}) &=& \int n(\textbf{k})dk_{z}.
\end{eqnarray}
Here, without loss of generality, we have assumed that the probe laser is applied along the $z$-direction.
With these, we calculate the column-integrated momentum distribution of the trapped lattice gas at various lattice depths across the critical point, from which we may extract various commonly used signatures for superfluidity.

We first plot in Fig.~1 the column-integrated momentum space density distribution along the $x$-axis
(with $k_{y}=0$) in the first Brillouin zone. Different from the previous studies, we fix the total number
$N=1.0 \times 10^{5}$ at a given temperature $T=0.18E_{\textrm{R}}$. The parameters are chosen
in close relation to existing experiments~\cite{mark-11}. From Fig.~1, we see that  bimodal structures emerge
as soon as the optical lattice depth is below the critical point. This is qualitatively consistent with the results
in Ref.~\cite{yi-07, lin-07}, where the total particle number is not fixed.

Another commonly used signature for the onset of superfluidity is the visibility of the interference
pattern~\cite{gerbier-05}
\begin{eqnarray}
  v &=&\frac{n^{A}_{\perp}-n^{B}_{\perp}}{n^{A}_{\perp}+n^{B}_{\perp}},
\end{eqnarray}
where $n^{A}_{\perp}$ and $n^{B}_{\perp}$ are column-integrated atomic intensities at site $A$ and $B$,
respectively. Here, point $A$ represents the position of a secondary peak while point $B$ is along a
diagonal with the same distance to the central peak as point $A$. Fig. 2 shows the visibility as a function of the scattering length, temperature, trapping frequency and the total particle number.
We find that the visibility monotonically decreases from unity
to a finite value by increasing the lattice depth. The finite visibility at temperatures above $T_c$
originates from the short-range correlations which are also present in a normal gas. However,
we notice that the visibility undergoes a sharp transition by crossing the critical optical lattice depth,
indicating its validity as a superfluid transition signature. We also observe that the value of this
residual visibility closely depends on the global harmonic trap and the total particle number, as
shown in Fig. 2(c) and 2(d).
\begin{figure}
\includegraphics[width=8.5cm]{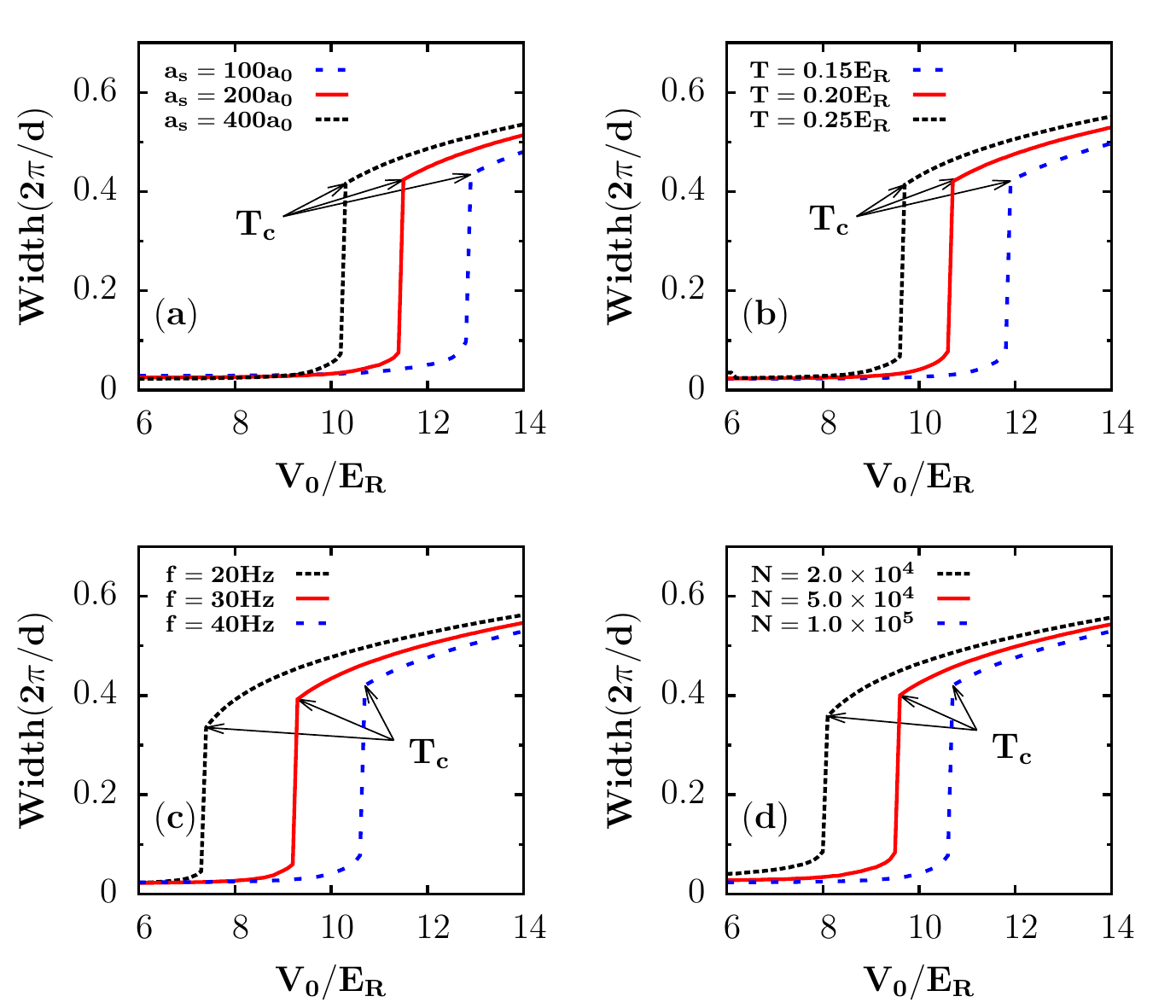}\\
\caption{(Color online) Peak width as a function of scattering length, temperature, trapping frequency and total particle number, respectively. The parameters are: (a) $T=0.2E_{\textrm{R}}$, $\omega=2\pi\times 40$ Hz and $N=1.0\times 10^{5}$; (b) $a_{s}=320a_{0}$, $\omega=2\pi\times 40$ Hz and $N=1.0\times 10^{5}$; (c) $a_{s}=320a_{0}$, $T=0.2E_{\textrm{R}}$ and $N=1.0\times 10^{5}$; (d) $a_{s}=320a_{0}$, $\omega=2\pi\times 40$ Hz and $T=0.2E_{\textrm{R}}$. }
\end{figure}


In order to characterize the sharpness of the atomic momentum distribution in the first Brillouin zone,
another commonly used single-value parameter is the peak width of the interference pattern,
which is defined as the half-peak width of $n_{\perp}(k_{x},k_{y})$ within the first Brillouin zone
along $k_{y}=0$
\begin{eqnarray}
  n_{\textrm{mid}} &\equiv&\frac{1}{2}\Big[n^{\textrm{max}}_{k_{x}}+n^{\textrm{min}}_{k_{x}}\Big].
\end{eqnarray}
In Fig.3 we show the peak width as a function of scattering length, temperature, trapping frequency
and the total particle number. Notice that the peak width increases monotonically with the lattice depth,
and undergoes a sharp change by crossing the critical transition point. Interestingly, the peak width
around the critical point typically features a two-platform structure, where we find that the second platform
at a higher lattice depth is associated with the superfluid to normal phase transition.
Indeed, the peak width features a sharp increase with an upward curvature when the system is
still in the superfluid region, and saturates when crossing the critical point. Thus, we suggest
that it is the second platform with saturating peak width that should be used as an unambiguous
signal for the phase transition, while the rising point which are used in existing experiments is still
within the superfluid region~\cite{mark-11}. Besides, we also notice that the saturating value of peak
width is also closely related to the global trapping potential and total particle number.



\section{Conclusion}\label{summary}
In summary, we have studied the finite-temperature properties of a trapped ultracold Bose gas
throughout the superfluid to normal phase transition, where the number of total particles is fixed.
Applying the HFBP formalism, we characterize various signatures associated with the column-integrated
momentum distribution of the lattice gas, which can be probed using typical time-of-flight imaging techniques.
From our calculations, we find that across the critical point, sharp features can be identified in all signatures
we considered, including the visibility and the peak width of the interference pattern.
In particular, we identify a two-platform structure in the width of the interference peak in the first Brillouin zone
as the lattice depth is tuned. We show that it is the higher platform in this two-platform structure that should
be used as a signature for the superfluid to normal phase transition.

The authors thank for support from NFRP (2011CB921200, 2011CBA00200), NKBRP (2013CB922000),
NNSF (60921091), NSFC (11105134, 11274009, 11374283), the Fundamental Research Funds
for the Central Universities (WK2470000006), and the Research Funds of
Renmin University of China (10XNL016, 14XNH061).

%
%

\end{document}